\documentclass[12pt,a4paper]{article}
\usepackage{axodraw}
\usepackage{cite}
\usepackage{epsfig}
\def\adhoc{\emph{ad hoc}}
\def\eg{\emph{e.g.}}
\def\ie{\emph{i.e.}}
\def\Z{\hphantom{-}}
\title{%
  \hfill\raisebox{2cm}[0pt][0pt]{\normalsize EPTCO-99-004} \\
  Phenomenology of $g_1(x)$ in \\
  the Observed Small-$x$ Region \\[6pt]
  \small (submitted to \emph{Phys.\ Lett.} \textbf{B})
}

\author{%
  Mehrdad Goshtasbpour $^{1,\,2,\,3,\,}$%
  \thanks{\,E-mail: \texttt{goshtasb@alborz.sbu.ac.ir}} \\
  and \\
  Philip G.~Ratcliffe $^{3,\,4,\,}$%
  \thanks{\,E-mail: \texttt{pgr@fis.unico.it}} \\[12pt]
  \hskip-1.5em
  \parbox{0.8\textwidth}{%
    \begin{list}
      {$^\arabic{enumi}$}
      {\small
       \topsep     0pt
       \partopsep  0pt
       \parsep     0pt
       \itemsep    0pt
       \labelwidth 1ex
       \labelsep   1pt
       \usecounter{enumi}%
       \def\makelabel##1{##1\hss}%
       \it}
      \item Dept.\ of Physics, Shahid Beheshti University,
      \\    Evin 19834, Tehran, Iran
      \item Center for Th.\ Phys.\ and Math., AEOI, Tehran, Iran
      \item Dip.\ di Scienze CC.FF.MM., Univ.\ degli Studi dell'Insubria,
      \\    via Lucini 3, 22100 Como, Italy
      \item Istituto Nazionale di Fisica Nucleare---sezione di Milano
    \end{list}}
}

\date{November 1999}
\begin{document}
\maketitle
%-----------------------------------abstract----------------------------------%
\begin{abstract}
  We examine the behaviour of the polarised structure function data
  for $g_1(x)$ in the region of small $x$ ($\sim0.01$ to $0.1$) for
  the various available targets (proton, neutron and deuteron) with
  the aim of clarifying what may be safely deduced with regard to the
  relation of the currently attained small-$x$ region and the
  asymptotic behaviour as $x\to0$. We find that fits using a single
  power-like term are susceptible to the isospin combination used.
  Double-power fits are more stable and also provide some evidence for
  an underlying SU(2) symmetry of the structure functions.
\end{abstract}
\newpage
%----------------------------------main body----------------------------------%
\section{Introduction}

As the data on polarised nucleon deep-inelastic structure functions
improves in precision, it becomes ever-more desirable to understand
the control that is possible over the asymptotic behaviour as $x\to0$
for fixed $Q^2$. This desire is driven by the necessity to extrapolate
to the point $x=0$ in order to compute (or estimate) integrals of the
structure functions (as required to test, \eg, the Bjorken sum rule).
Indeed, while it is generally held that such integrals are convergent,
a steeply rising behaviour of the type $x^{-0.9}$, for example, would
leave a significant part of the related integral in the unmeasured
region and could invalidate any deductions derived therefrom.

The problem has already been examined\cite{Soffer:1997ft, Bass:1997fh}
using varying degrees of model-dependent analysis. In this letter we
shall attempt to analyse the situation starting from the least
model-dependent approach and then investigate the effects of various
possible assumptions. In particular, we shall examine to what extent
simultaneous double-power fits of all the data provide a more coherent
picture as compared to combinations of single-power fits to the single
data sets; and what indications there exist for an underlying isospin
structure.

We shall avoid any direct reference to Regge theory (save that the
structure functions should behave as powers of $x$ asymptotically),
the effect of including model input from Regge-based analyses has been
discussed in \cite{Bass:1997fh}. We note in passing that
in\cite{Bass:1997fh} the basic input was a logarithmic behaviour of
the leading isoscalar contribution.  However, given the very short
lever-arm available in $x$ (the SLAC data between about $x=0.01$ and
$0.15$ were used), any logarithmic variation can easily be
approximately reproduced by a power behaviour and is therefore not to
be considered fundamental at this stage of the analysis.

The layout of this letter is as follows: in Section~\ref{sec:1pfit} we
discuss the stability of single-power fits to differing combinations
of the data, in Section~\ref{sec:2pfit} the analysis is extended to
include double-power contributions and in Section~\ref{sec:isofit} the
effect of isospin assumptions is examined. We end with a series of
conclusions and comments.

%-----------------------------------section-----------------------------------%
\section{Single-Power Fits}
\label{sec:1pfit}

In\cite{Soffer:1997ft} an effectively single-power fit for $g_1^n$ was
performed, with the conclusion that the dominant power behaviour was
of the form
\begin{equation}
  g_1^n \simeq -0.015 x^{-0.87}.
  \label{eq:g1n}
\end{equation}
If such were the case, the unmeasured region, below $x\simeq0.01$ say,
would contribute ${\sim}0.06$ to the $g_1^n$ integral. Since even the
region below $x\simeq0.001$ would still contribute $\sim0.05$, this
would render any future test of the integral on very unsure footing.
To save the Bjorken integral from unwanted implications, in
\cite{Soffer:1997ft} the authors were careful to estimate the triplet
$g_1^p(x)-g_1^n(x)$ combination independently with a smaller,
negative, power term $\sim-0.45$.

\begin{figure}[htbp]
  \begin{center}
    \epsfbox{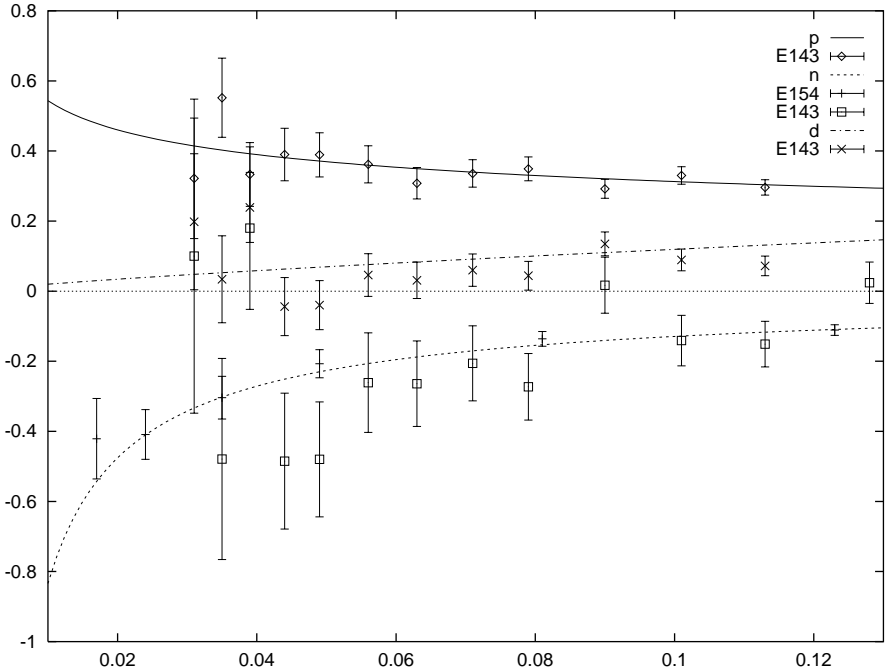}
    \caption{Comparison of the single-power fitted curves and the data.}
    \label{fig:1pfit}
  \end{center}
\end{figure}

Thus, to provide a starting point, we first perform individual fits to
the SLAC E143 data at $5\,$GeV$^2$ for the three experimental targets:
proton, neutron and deuteron\cite{Abe:1998wq} using a single-power
contribution of the form
\begin{equation}
  g_1 = \alpha x^\delta,
  \label{eq:1pfit}
\end{equation}
where $\alpha$ and $\delta$ are then free parameters. The results are
summarised in Table~\ref{tab:1pfit}, where we also include the
hypothetical isovector target $p-n$. All fits return perfectly
acceptable values of $\chi^2$ although the powers found differ widely.
A comparison of the fitted curves and the data is presented in
Fig.~\ref{fig:1pfit}
\begin{table}[htb]
  \begin{center}
    \caption{Results for single-power fits to the SLAC E143 proton, neutron
      and deuteron data, at $5\,$GeV$^2$ as described in the text.}
    \vspace{1ex} $\begin{array}{ccc} \hline\hline
      \mbox{target} &   \alpha     &    \delta    \\
      \hline
      g_1^p         &\Z0.18\pm0.05 & -0.24\pm0.11 \\
      g_1^n         & -0.02\pm0.02 & -0.81\pm0.40 \\
      g_1^d         &\Z0.72\pm1.05 &\Z0.78\pm0.61 \\
      g_1^p-g_1^n   &\Z0.10\pm0.05 & -0.51\pm0.18 \\
      \hline\hline
    \end{array}$
    \label{tab:1pfit}
  \end{center}
\end{table}

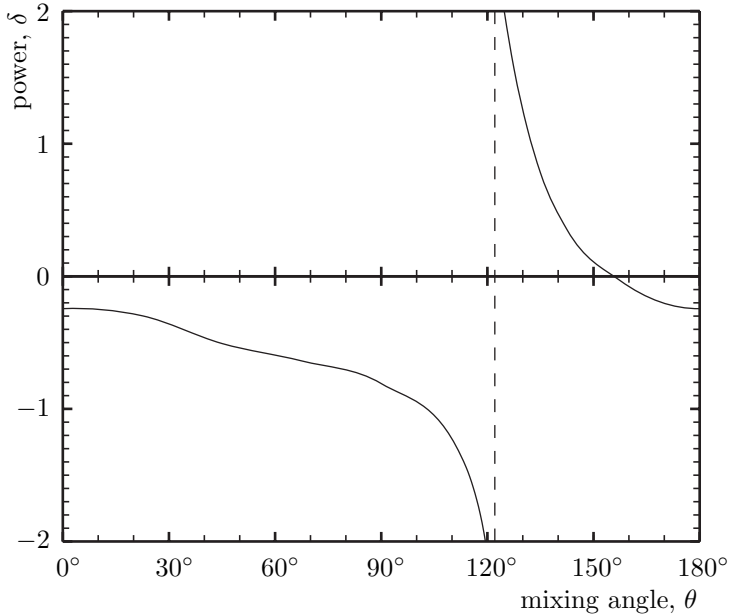
\begin{figure}[htbp]
  \centering %==============================================================================%
% theta.tex ((La)TeX input)                                                    %
% last modified on 30-Aug-1999                             Philip G. Ratcliffe %
%==============================================================================%
% An axodraw graphic picture
%==============================================================================%
\ifx\ZigZag\undefined
  \GenericError{}
               {The above picture file requires Vermaseren's "axodraw" package}
               {The picture will be ignored.^^J}
               {Insert \protect\usepackage{axodraw} in the preamble.}
   
\fi
\bgroup
\footnotesize
\begin{picture} (280, 240)(-20,-120)
% horizontal axes
\LinAxis        (  0,-100)(240,-100)( 6, 3, 3, 0, 1)
\LinAxis        (  0, 100)(240, 100)( 6, 3,-3, 0, 1)
\LinAxis        (  0,   0)(240,   0)( 6, 3, 3, 0, 1)
\LinAxis        (  0,   0)(240,   0)( 6, 3,-3, 0, 1)
% vertical axes
\LinAxis        (  0,-100)(  0, 100)( 4,10,-3, 0, 1)
\LinAxis        (240,-100)(240, 100)( 4,10, 3, 0, 1)
\rText          (-18, 100)[t][l]{power, $\delta$}
\Text           ( -5, 100)[r]{$ 2$}
\Text           ( -5,  50)[r]{$ 1$}
\Text           ( -5,   0)[r]{$ 0$}
\Text           ( -5, -50)[r]{$-1$}
\Text           ( -5,-100)[r]{$-2$}
\Text           (240,-123)[r]{mixing angle, $\theta$}
\Text           (  0,-110)[ ]{$  0$\rlap{$^\circ$}}
\Text           ( 40,-110)[ ]{$ 30$\rlap{$^\circ$}}
\Text           ( 80,-110)[ ]{$ 60$\rlap{$^\circ$}}
\Text           (120,-110)[ ]{$ 90$\rlap{$^\circ$}}
\Text           (160,-110)[ ]{$120$\rlap{$^\circ$}}
\Text           (200,-110)[ ]{$150$\rlap{$^\circ$}}
\Text           (240,-110)[ ]{$180$\rlap{$^\circ$}}
% Here is the curve:
% to the left of the discontinuity
\SetScale{50}\SetWidth{0.01}
\Curve{%  power     theta
  (0.000,-0.244) %   0.00
  (0.600,-0.298) %  22.50
  (1.200,-0.507) %  45.00
  (1.800,-0.638) %  67.50
  (2.400,-0.809) %  90.00
  (3.000,-1.353) % 112.50
  (3.040,-1.441) % 114.00
  (3.067,-1.512) % 115.00
  (3.093,-1.592) % 116.00
  (3.120,-1.687) % 117.00
  (3.147,-1.798) % 118.00
  (3.173,-1.929) % 119.00
  (3.185,-2.000) % 119.44 % interpolated
% (3.200,-2.088) % 120.00
% (3.227,-2.292) % 121.00
}
% the discontinuity
\DashLine(3.256,-2.000)(3.256, 2.000){.1}
% to the right of the discontinuity
\Curve{%  power     theta
% (3.300, 2.159) % 123.75
  (3.327, 2.000) % 124.76 % interpolated
  (3.400, 1.569) % 127.50
  (3.500, 1.119) % 131.25
  (3.600, 0.780) % 135.00
  (3.800, 0.352) % 142.50
  (4.200,-0.032) % 157.50
  (4.800,-0.244) % 180.00
}
\end{picture}
\egroup
%==============================================================================%
 
  \caption{The power, $\delta$, obtained in a single-power fit as a function
    of $\theta$, the mixing parameter for $g_1^p$ and $g_1^n$
    described in the text.}
  \label{fig:g1theta}
\end{figure}

As is clear from the results displayed in Table~\ref{tab:1pfit}, the
precise power returned in a single-power fit is highly dependent on
the target combination used for the data, varying between $-0.81$ for
the neutron and $0.78$ for the deuteron. Thus, it is not possible,
even in a first approximation to assume one universal power to
describe this region of $x$.  While this is clearly an indication of
the necessity to use at least two powers, in view of the results of,
\eg,\cite{Soffer:1997ft}, where the neutron emerged well described by
a single power (as too here), it will be instructive to study the
variation of the power returned as a function of the target
considered. To this aim we perform an analysis by fitting single
powers to combinations of the data according to the formula
\begin{equation}
  g_1^\theta = \sin\theta\,g_1^n - \cos\theta\,g_1^p
  \label{eq:g1theta}
\end{equation}
and vary the parameter $\theta$ from $0$ to $180^\circ$, thus covering
the full range including both $g_1^d$ and the combination
$g_1^p-g_1^n$. The results are displayed in Fig.~\ref{fig:g1theta}.
The obvious feature is the sharp discontinuity located at
$\theta\sim120^\circ$, which clearly indicates the failure of a
single-power fit in that region of combinations, very close to the
deuteron. What should also not be underestimated is the tail of this
discontinuity, which extends into the region of pure $g_1^n$.

As commented near the end of the paper, the few (2 or 3) smallest-$x$
data points (below $x\sim0.03$) of the E155\cite{Anthony:1999rm} (and
SMC) experiments (at $5\,$GeV$^2$) hint that more small-$x$ data in
the future may push the deuteron fitted curves towards negative
powers, which on the $\theta$ plot of a single-power fit would mean a
discontinuity and thus inconsistency.

A consistent set of fits, related to the procedure
of\cite{Soffer:1997ft}, is provided by taking the functions fitted to
$g_1^n$ and $g_1^p-g_1^n$ as ``primary'' and obtaining $g_1^p$ and
$g_1^d$ as linear combinations, so that there are only two independent
powers (all functions have acceptable $\chi^2$). As the starting point
may be any two of the four functions (with single powers), there is to
be an element of arbitrariness here. Such arbitrariness is implicitly
exploited in\cite{Soffer:1997ft} where the single-power fit to $g_1^n$
is considered more fundamental or ``primary'', thus leading to a very
divergent $g_1^n$. However, such behaviour disappears if $g_1^n$ is
taken as a linear combination, as also intuitively suggested
in\cite{Ratcliffe:1996nc}. Is there anything then that favours two of
the functions and their associated single-powers as more fundamental
than the other two?

Here, we are led once again to consider two separate power-term
contributions, at least in two of the four combination $g_1$ functions
considered. Thus, it is reasonable to begin with a two-power global
fit to the structure functions to avoid all arbitrariness. This ought
to be done simultaneously for all the available data. We actually used
SLAC and HERMES at approximately $3\,$GeV$^2$.

Before moving on to double-power fits, it is worth investigating the
possibility of describing both the proton and neutron data with the
same single power (\ie, explicitly leaving aside the deuteron data for
the moment); the results are
\begin{equation}
  \begin{array}{rcl}
     \alpha_p & = &\Z0.10\pm0.02 \\
     \alpha_n & = & -0.05\pm0.01 \\
     \delta\; & = & -0.48\pm0.08.
  \end{array}
  \label{eq:pnpower}
\end{equation}
However, the $\chi^2$ here is significantly poorer than in any other
fit so far. On examining these results together with the data, it
emerges that the poor fit could be readily remedied by the inclusion
of a plateau increasing the overall proton polarisation and decreasing
that of the neutron. The effect would then be to reduce the neutron
power while increasing that of the proton, leaving a slightly positive
deuteron (given by the plateau contribution itself).

%-----------------------------------section-----------------------------------%
\section{Double-Power Fits}
\label{sec:2pfit}

Not only is it obviously necessary to allow for more than one power in
the fits, but it should also be clear that any single power returned
above could change dramatically in a two-power
fit\cite{Ratcliffe:1996nc}. In particular, two facts lead to the
possibility that any given behaviour might be mimicked by combinations
of different powers.

First of all is the limited lever-arm available in terms of the
$x$-range of the data one can sensibly use; little more than one order
of magnitude variation in $x$ is the bare minimum for extracting a
power behaviour, given the present experimental errors. Second, and
often neglected, is the fact that, in contrast to the unpolarised
case, where most of the standard wisdom has been gathered, the
polarised structure functions are \emph{not} constrained to be
positive definite. This means that, \eg, an overall rising behaviour
(in magnitude) may be due either to a single rising contribution
\emph{or} to a rising cancellation between two relatively changing
contributions of opposite sign, neither of which need necessarily be
similarly rising.

In\cite{Ratcliffe:1996nc} an explicit example was invented \adhoc, in
which a behaviour of the type (as proposed in\cite{Soffer:1997ft})
\begin{equation}
  g_1^n(x) \sim \frac{-0.02}{x^{0.8}}
  \label{eq:stpower}
\end{equation}
was shown to be well reproduced, within experimental errors and over
the finite range of $x$ considered, with the form
\begin{equation}
  g_1^n(x) \sim \frac{-0.07}{x^{0.5}}(1-4x),
  \label{eq:pgrpower}
\end{equation}
which, one should note, has very different asymptotic behaviour.

\begin{figure}[htb]
  \begin{center}
    \epsfbox{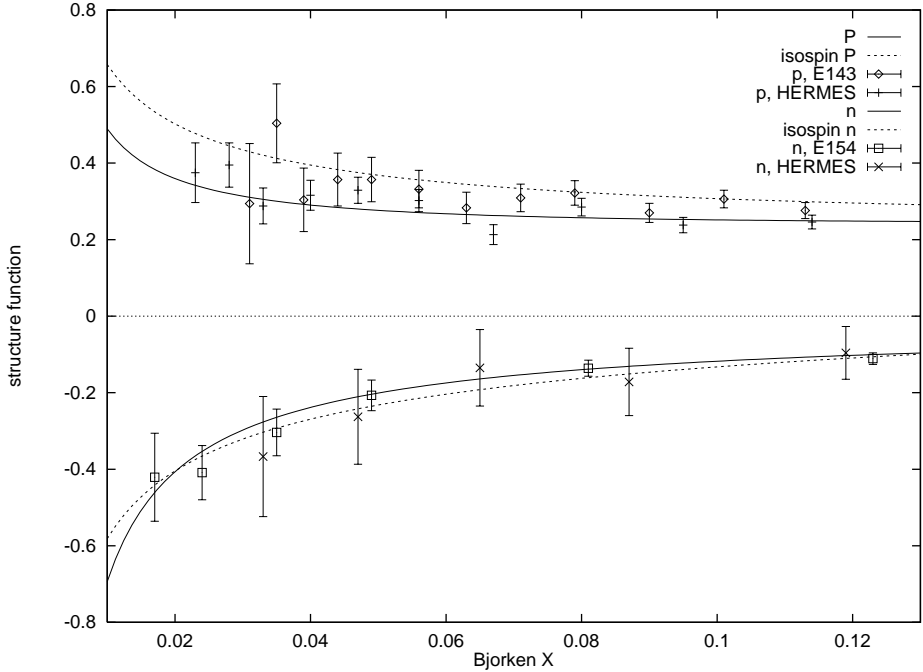}
    \caption{Comparison of the double-power fitted curves (solid lines) and
      isospin type fits (dashed lines) with the data.}
    \label{fig:2pfit}
  \end{center}
\end{figure}

The fact that single-power fits do, in fact, work rather well implies
that an attempt to fit any single target data set (excepting perhaps
that of the deuteron) with two power terms will encounter serious
difficulties. Indeed, the only way to perform successful two-power
fits is to combine data sets of different targets and demand that the
two powers used be the same. The results of such a fit can be
summarised as follows: for the form
\begin{equation}
  g_1 = \alpha x^\delta + \beta x^\gamma,
  \label{eq:2pform}
\end{equation}
and the SLAC\cite{Abe:1998wq,Abe:1997dp} E143, E154 together with the
HERMES\cite{Ackerstaff:1997ws} data sets, we obtain
\begin{equation}
  \begin{array}{rcl}
    \alpha_p & = &\Z0.01 \pm 0.02 \\
    \alpha_n & = & -0.02 \pm 0.03 \\
    \delta\; & = & -0.77 \pm 0.33 \\
    \beta_p  & = &\Z0.26 \pm 0.20 \\
    \beta_n  & = &\Z0.00 \pm 0.10 \\
    \gamma\; & = &\Z0.13 \pm 0.50
  \end{array}
  \label{eq:2pfit1}
\end{equation}
and again the $\chi^2$ is perfectly acceptable. Note that the large
errors are mainly due to strong correlations between coefficients and
powers and therefore ultimately between the coefficients themselves.
Again, a comparison of the fits and data is shown in
Fig.~\ref{fig:2pfit}.

One sees the attraction of the neutron data to a single-power fit,
\ie, that a second power is not at all required by this particular
target. Moreover, the fact that the power chosen is very close to the
original power but far from that of the proton is indicative of the
possibility to mimic powers (in this case that of the proton) by
combining different contributions. Note also that, as has been found
by various authors, the steeper behaviour ($\delta$ here), having
opposite sign coefficients ($\alpha$) in the proton and neutron, is to
be ascribed to an isovector contribution---within errors the two
coefficients have the same magnitude.

As a final test we fix the steep power to be less divergent, setting
$\delta=-0.5$, and repeat the fits---the idea being to present a
picture less clouded by the correlations between parameters, which
often artificially inflate errors; the results are
\begin{equation}
  \begin{array}{rcl}
    \alpha_p & = &\Z0.03 \pm 0.05 \\
    \alpha_n & = & -0.07 \pm 0.03 \\
    \delta\; & = & -0.50 \quad\mbox{fixed} \\
    \beta_p  & = &\Z0.17 \pm 0.09 \\
    \beta_n  & = &\Z0.11 \pm 0.07 \\
    \gamma\; & = &\Z0.02 \pm 0.63
  \end{array}
  \label{eq:2pfit2}
\end{equation}
as always the $\chi^2$ is perfectly acceptable and indeed marginally
improves due to the reduced number of parameters.

Hence it is clear that such steep powers as previously found are not
an absolute requirement of the data, as suggested by the large errors
found earlier. Moreover, on choosing a less divergent behaviour the
various coefficients come more into line with isospin symmetry
requirements; \ie, one term is approximately isovector and the other
isoscalar.

%-----------------------------------section-----------------------------------%
\section{Isospin Symmetric Fits}
\label{sec:isofit}

In the light of the above fits, it is natural to enquire as to the
effect of requiring that the two terms used fall precisely into the
two categories of isovector and isoscalar. Thus, as a final test we
have simply repeated the above fit fixing $\alpha_n=-\alpha_p$ and
$\beta_n=\beta_p$. The results are (with the usual good $\chi^2$)
\begin{equation}
  \begin{array}{rcl}
    \alpha_p & = &\Z0.78 \pm 0.02 \\
    \delta\; & = & -0.45 \pm 0.07 \\
    \beta_p  & = &\Z0.20 \pm 0.14 \\
    \gamma\; & = &\Z0.36 \pm 0.28 ,
  \end{array}
  \label{eq:isofit}
\end{equation}
where $\delta$ and $\gamma$ now refer to the isovector and isoscalar
powers respectively.

Relaxing one or other of the isospin constraints leads to similar fits
with similar $\chi^2$. Note that in this case the steep power, which
is, of course, now constrained to be that of the combination
$g_1^p-g_1^n$ is much less steep than in many other fits. This would
mean an increasing difference in total polarisation of $u$ and $d$
quarks, with decreasing $x$ in the given interval.  A comparison of
the fits and data can be found in Fig.~\ref{fig:2pfit}.

Thus, we may say that in the small-$x$ region under examination, the
double-power fits of the data prior to the E155 experiments are
consistent with the assumption that the positive isotriplet
contribution is causing the mildly divergent behaviour of $g_1^p(x)$
and $g_1^n(x)$. However, the latest E155 data on the
deutron\cite{Anthony:1999rm}, which are still preliminary, provide a
hint of a negative singlet term in $g_1^d(x)$ with a more strongly
divergent behaviour, setting in below $x=0.03$, in accordance with the
expected\cite{Abe:1997dp} asymptotic dominance of the singlet. This,
together with the observed attraction of the neutron data to a single
steep power, calls for a fresh look at the picture when the final E155
data become available.

%-----------------------------------section-----------------------------------%
\section{Conclusions}

The first lesson that we wish to underline is the possibility that any
result arising from an effectively \emph{single}-power fit may be
biased precisely by the choice of a single term to describe the data.
Thus, although the neutron data appears to select a large single
power, two lesser powers are also perfectly acceptable (and less
dramatic). Indeed, by comparing the magnitudes of the isoscalar and
isovector contributions, one sees that over much of the $x$ range of
interest neither contribution is negligible. It is also clear that the
presence of an isoscalar plateau leads to a steepening of one power
and a flattening of the other.

Perhaps, the second lesson is that, while the isospin properties of
the distributions are still far from well defined in the data, a
simple description based on such a symmetry does, in fact, work rather
well.

One caveat that should be borne in mind is that we have not considered
how evolution might alter the picture: this is both beyond the scope
of the present letter and, to any reliable degree, beyond the level of
the present data given the poor determination of the gluon
distribution.
%-------------------------------acknowledgments-------------------------------%
\section{Acknowledgments}

One of us (M.G.) is grateful to the Department of Sciences, University
of Insubria at Como, for hospitality while this analysis was performed
and is also grateful to the Shahid Beheshti University for funded
sabbatical leave during this research.
%---------------------------------bibliography--------------------------------%
%\pigrobib
%\end{document}
%%%%%%%%%%%%%%%%%%%%%%%%%%%%%%%%%%%%%%%%%%%%%%%%%%%%%%%%%%%%%%%%%%%%%%%%%%%%%%%
% DO NOT ADD ANY NEW ITEMS BELOW THIS LINE; THE BIBLIOGRAPHY IS GENERATED     %
% AUTOMATICALLY.  NEW CITATIONS SHOULD BE INDICATED IN THE TEXT AND CAN       %
% THEN BE ADDED VIA BIBTEX LATER.                                             %
%%%%%%%%%%%%%%%%%%%%%%%%%%%%%%%%%%%%%%%%%%%%%%%%%%%%%%%%%%%%%%%%%%%%%%%%%%%%%%%

%------------------------------------the end----------------------------------%
\end{document}